\documentclass[twocolumn,english,prb,superscriptaddress]{revtex4-1}
\usepackage[T1]{fontenc}
\usepackage[latin9]{inputenc}
\usepackage{units}
\usepackage{amsmath}
\usepackage{amssymb}
\usepackage{graphicx}
\usepackage{esint}

\makeatletter

\providecommand{\tabularnewline}{\\}

\usepackage{babel}

\makeatother

\usepackage{babel}
\begin{document}

\title{Evolution of coherent collective modes through consecutive CDW transitions
in (PO$_{2}$)$_{4}$(WO$_{3}$)$_{12}$ mono-phosphate tungsten bronze}

\author{L. Stojchevska}

\address{Complex Matter Department, Jozef Stefan Institute, Jamova 39, 1000
Ljubljana, Slovenia}

\author{M. Borovšak}

\address{Complex Matter Department, Jozef Stefan Institute, Jamova 39, 1000
Ljubljana, Slovenia}

\address{Faculty of Mathematics and Physics, University of Ljubljana, Jadranska
19, 1000 Ljubljana, Slovenia}

\author{P. Foury-Leylekian}

\address{Laboratoire de Physique des Solides, UMR CNRS 8502, Université Paris
Sud, 91405 Orsay - France}

\author{J.-P. Pouget}

\address{Laboratoire de Physique des Solides, UMR CNRS 8502, Université Paris
Sud, 91405 Orsay - France}

\author{T. Mertelj}

\email{tomaz.mertelj@ijs.si}

\selectlanguage{english}%

\address{Complex Matter Department, Jozef Stefan Institute, Jamova 39, 1000
Ljubljana, Slovenia}

\address{Center of Excellence on Nanoscience and Nanotechnology Nanocenter
(CENN Nanocenter), Jamova 39, 1000 Ljubljana, Slovenia}

\author{D. Mihailovic}

\address{Complex Matter Department, Jozef Stefan Institute, Jamova 39, 1000
Ljubljana, Slovenia}

\address{Faculty of Mathematics and Physics, University of Ljubljana, Jadranska
19, 1000 Ljubljana, Slovenia}

\address{Center of Excellence on Nanoscience and Nanotechnology Nanocenter
(CENN Nanocenter), Jamova 39, 1000 Ljubljana, Slovenia}

\date{\today}
\begin{abstract}
All optical femtosecond relaxation dynamics in a single crystal of
mono-phosphate tungsten bronze (PO$_{2}$)$_{4}$(WO$_{3}$)$_{2m}$
with alternate stacking $m=6$ of WO$_{3}$ layers was studied through
the three consequent charge density wave (CDW) transitions. Several
transient coherent collective modes associated to the different CDW
transitions were observed and analyzed in the framework of the time
dependent Ginzburg-Landau theory. Remarkably, the interference of
the modes leads to an \emph{apparent} \emph{rectification} effect
in the transient reflectivity response. A saturation of the coherent-mode
amplitudes with increasing pump fluence well below the CDWs destruction
threshold fluence indicates a decoupling of the electronic and lattice
parts of the order parameter under strong optical drive.
\end{abstract}
\maketitle

\section{Introduction}

Time-resolved spectroscopy can give unique insight in the dynamical
behaviour of the elementary excitations in systems undergoing charge
density wave (CDW) transitions. The appearance of new collective vibrational
excitations in Raman\cite{travaglini1983,Sugai1985,lavagnini2010}
and coherent transient reflectivity\cite{DemsarBlueBronze,Shimtake2007,YusupovPRL101}
response upon charge density wave (CDW) formation is well documented\cite{travaglini1983,Sugai1985,DemsarBlueBronze,Shimtake2007,lavagnini2010,Schaefer2010,YusupovPRL101}
and theoretically quite well understood.\cite{Schaefer2010} In systems
with more than one CDW transition new collective modes are expected
to appear below each transition temperature, as observed experimentally
in the Raman response\cite{Sugai1985,lavagnini2010}. In the transient
reflectivity response\cite{YusupovPRL101,PRBMilos}, however, the
additional coherent oscillation modes below the subsequent CDW transition
temperatures have not been readily observed. Investigating the differences
in the response associated with different transitions gives important
information on the coupling and damping of collective modes, and behaviour
of the order parameter in each case.

\begin{figure}[h!]
\includegraphics[angle=-90,width=1\columnwidth]{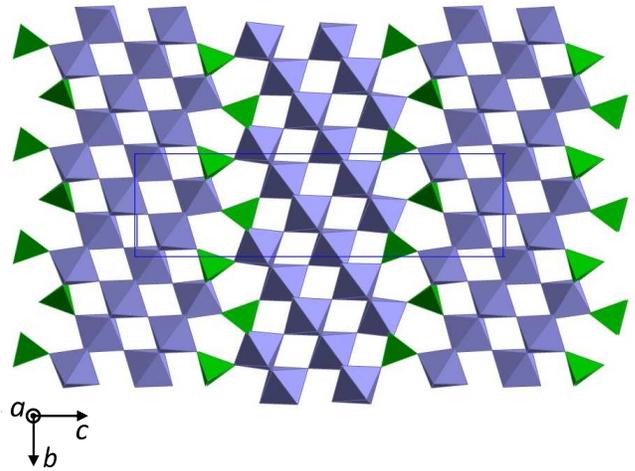}
\caption{Crystal structure of (PO$_{2}$)$_{4}$(WO$_{3}$)$_{12}$ showing
the WO$_{6}$ octahedra and the mono-phosphate PO$_{4}$ tetrahedra
projected along the $a$ crystallographic direction.}
\label{fig:Figure1} 
\end{figure}

The mono-phosphate tungsten bronzes (PO$_{2}$)$_{4}$(WO$_{3}$)$_{2m}$
are an example of low dimensional CDW materials\cite{FouryEPL} offering
a possibility to further explore the effect of consecutive CDW transitions
on the transient-reflectivity coherent oscillatory response. They
are quasi two-dimensional (2D) conductors built from layers of WO$_{6}$
octahedra parallel to the $(a,b)$-plane and separated by PO$_{4}$
tetrahedra \cite{IJMPPouget93,Roussel,FourySolidState,SlobodynikRussian}
as shown in Fig. \ref{fig:Figure1}. Their electronic properties arise
from quasi-one-dimensional chain structures where - due to the strong
Fermi surface nesting - charge-density-wave (CDW) instabilities develop
below certain critical transition temperatures $T_{\mathrm{CDW}}$,
as confirmed by various experimental techniques\cite{IJMPPouget93,SchlenkerJPhys96,Greenblatt}.
For (PO$_{2}$)$_{4}$(WO$_{3}$)$_{2m}$ with $m=6$ in particular
three CDW transitions have been observed: $T_{\mathrm{CDW1}}=120$~K,
$T_{\mathrm{CDW2}}=62$~K and $T_{\mathrm{CDW3}}=30$~K\cite{IJMPPouget93,WangPRB,FouryEPL}.

Up to date, a substantial effort has been put into studying the electronic,
magnetic and structural properties of (PO$_{2}$)$_{4}$(WO$_{3}$)$_{2m}$
bronzes, \cite{Schlenker1,Greenblatt1,Schlenker2} however, to the
best of our knowledge there is no Raman or time-resolved optical spectroscopy
study of the sequence of CDW phases in (PO$_{2}$)$_{4}$(WO$_{3}$)$_{2m}$
tungsten bronzes. In this paper we therefore report on temperature
and fluence dependence of the transient reflectivity in a mono-phosphate
tungsten bronze (PO$_{2}$)$_{4}$(WO$_{3}$)$_{2m}$ single crystal
with $m=6$ (P$_{4}$W$_{12}$O$_{44}$) on a femtosecond time scale,
focusing on the effect of the subsequent CDW transitions on the coherent
oscillatory transient response.

\section{Experimental}

Single crystals used in this study were grown at CRISMAT (Caen, France)
according to a previously reported method.\cite{Labbe} They have
a shape of platelets with large surface of about (1$\times$1.5) mm$^{2}$,
corresponding to the $(a,b)$ conducting plane, and a thickness of
$\sim$ 1/10~mm along c. The index $m=6$ of the bronze was unambiguously
determined by the measurement of the inter-slab periodicity \cite{Roussel}
$c$ ($c=23.57$~$\AA$ for $m=6$).

Optical experiments were performed with 50~fs laser pulses at 800~nm
generated from an amplified Ti:Sapphire mode-locked laser at a 250~kHz
repetition rate. The transient reflectivity $\Delta$$R/R=\frac{R_{p}-R_{0}}{R_{0}}$,
where $R_{p}$ and $R_{0}$ are the reflectivities in the absence
and presence of the pump pulse, respectively, was monitored using
a standard pump-probe technique where both pump and probe photons
were at 1.55~eV photon energy. The beam diameters on the sample were
determined by means of calibrated pinholes to be 108~$\mu$m and
50~$\mu$m for the pump and probe beam, respectively. Both beams
were perpendicularly polarized to each other and oriented relatively
to the crystal axes in a way to obtain maximum/minimum response amplitude
at low temperatures. Before optical measurements the single crystals
were cleaved by means of scotch tape and mounted on a cold finger
of a liquid-He flow optical cryostat equipped with CaF$_{2}$ windows.

\section{Results}

\subsection{Experimental data}

\begin{figure}[h!]
\includegraphics[angle=270,width=1\columnwidth]{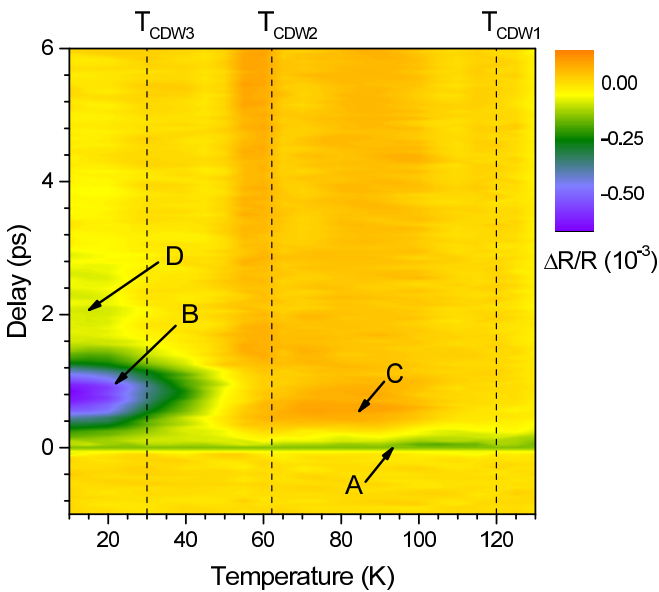} \caption{Temperature dependence of the transient reflectivity in (PO$_{2}$)$_{4}$(WO$_{3}$)$_{12}$
measured at the 8~$\mu$J/cm$^{2}$ pump fluence. Vertical dashed
lines indicate the transition temperatures: $T_{\mathrm{CDW3}}=30$~K,
$T_{\mathrm{CDW2}}=62$~K and $T_{\mathrm{CDW1}}=120$~K.}
\label{fig:Figure2} 
\end{figure}

\begin{figure}[h!]
\includegraphics[angle=270,width=0.8\columnwidth]{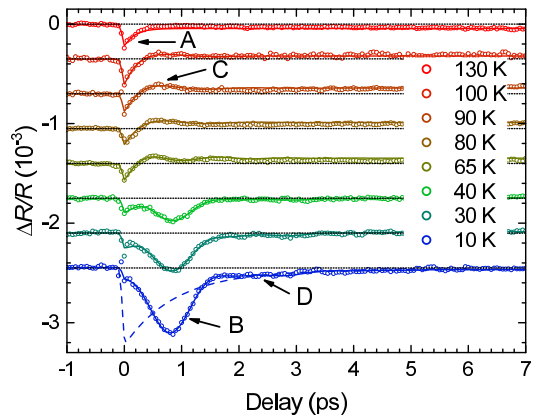} \caption{Transient reflectivity in (PO$_{2}$)$_{4}$(WO$_{3}$)$_{12}$ at
selected temperatures recorded with 8~$\mu$J/cm$^{2}$ pump fluence.
The transients are vertically shifted for clarity with the dotted
lines indicating the shifts. The continuous lines are DCE-model fits
discussed in text. The dashed line corresponds to the the displaced
equilibrium position for oscillators at 10 K. There is a notable coherent
artifact due to the pump scattering appearing as increased noise at
zero delay.}
\label{fig:Figure3} 
\end{figure}

\begin{figure}[h!]
\includegraphics[angle=270,width=0.8\columnwidth]{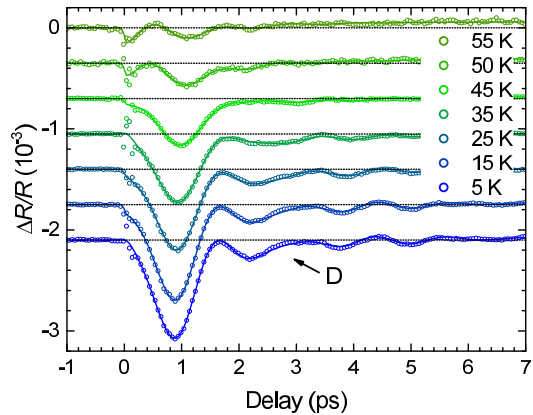}
\caption{Transient reflectivity in (PO$_{2}$)$_{4}$(WO$_{3}$)$_{12}$ in
the low-$T$ region recorded with 8~$\mu$J/cm$^{2}$ pump fluence.
The transients are vertically shifted for clarity with the dotted
lines indicating the shifts. The continuous lines are DCE-model fits
discussed in text. There is a notable coherent artifact due to the
pump scattering appearing as increased noise at zero delay.}
\label{fig:DRvsTlow} 
\end{figure}

In Figs. \ref{fig:Figure2}, \ref{fig:Figure3} and \ref{fig:DRvsTlow}
we show the temperature dependence of the transient reflectivity ($\Delta$$R/R$)
in (PO$_{2}$)$_{4}$(WO$_{3}$)$_{12}$ taken at a pump fluence of
8~$\mu$J/cm$^{2}$. The response is characterized by four main features. 

The first is a sub-picosecond negative transient (feature A), which
changes only little with decreasing $T$. 

The second and the most prominent is a larger picosecond negative
transient (feature B) that appears on top of feature A below $T_{\mathrm{CDW2}}=62$~K.
Remarkably, feature B (Fig. \ref{fig:Figure3}) \emph{displays a risetime
that appears slightly longer than the decay time}.

The third is a small positive feature C (see Fig. \ref{fig:Figure3})
peaked at $\sim0.5$ ps that appears below $T_{\mathrm{CDW1}}=120$~K
and develops into a weak oscillation when approaching $T_{\mathrm{CDW2}}$. 

The fourth is a negative shoulder together with additional weak oscillations
beyond $\sim2$ ps delay (feature D in Figs. \ref{fig:Figure3} and
\ref{fig:DRvsTlow}), that appear below the lowest transition at $T_{\mathrm{CDW3}}=30$~K. 

The feature-D weak oscillations were not completely reproducible between
different experimental runs. While the $T$-dependence shown in Fig.
\ref{fig:DRvsTlow} was measured in the first run immediately after
cleaving the sample the data shown in Fig. \ref{fig:Figure3} were
measured in the second run after keeping the sample in vacuum for
12 days. Since it is not possible to measure exactly the same sample
spot in different runs it is not clear whether the difference can
be attributed to deterioration of the sample surface or surface inhomogeneity.

\begin{figure}
\includegraphics[clip,angle=270,width=0.9\linewidth]{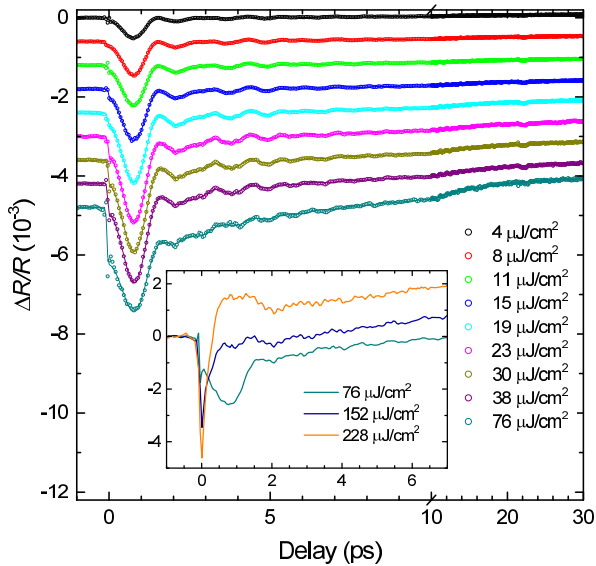}\caption{Pump fluence dependence of the transient reflectivity at 5 K. The
traces are vertically shifted for clarity. The thin lines are the
DCE-model fits discussed in text. The inset shows the qualitative
change at the fluences above $76$ $\mu$J/cm$^{2}$. }
\label{fig:DRvsF} 
\end{figure}

In Fig. \ref{fig:DRvsF} we plot the pump fluence dependence of the
transient reflectivity measured at $T=5$ K during the first run.
The magnitude of the transients shows saturation with increasing $\mathcal{F}$
above $\sim20$ $\mu$J/cm$^{2}$. With further fluence increase the
response changes qualitatively above $\mathcal{F}\simeq80$ $\mu$J/cm$^{2}$
as shown in the inset of Fig. \ref{fig:DRvsF}.

\subsection{Analysis}

An unusual feature of the data, which we need to understand, is that
at the lowest temperatures the risetime of feature B apparently exceeds
the decay time. Due to the ubiquitous appearance of a coherent collective
oscillations in the transient reflectivity upon CDW formation\cite{DemsarBlueBronze,Shimtake2007,YusupovPRL101}
it is very likely that the unusual shape of feature B is due to an
interference between coherent oscillations. To test this hypothesis
and get better insight into the unusual shape of feature B we fit
the data using the theory of displacive coherent excitation (DCE)
of oscillatory modes\cite{zeiger1992theory}: 
\begin{gather}
\frac{\Delta R}{R}=(A\mathrm{_{displ}}-\sum A_{\mathrm{O}i})\int_{0}^{\infty}G(t-u)e^{-u/\tau_{\mathrm{displ}}}du+\nonumber \\
+\sum A_{\mathrm{O}i}\int_{0}^{\infty}G(t-u)e^{-\gamma_{i}u}[\cos(\Omega_{i}u)-\nonumber \\
-\beta_{i}\sin(\Omega_{i}u)]du+\nonumber \\
+\sum_{j\in\{1,2\}}A_{\mathrm{e}j}\int_{0}^{\infty}G(t-u)e^{-u/\tau_{j}}du,\label{eq:fitfunc}
\end{gather}
with $\beta_{i}=(1/\tau_{\mathrm{displ}}-\gamma_{i})/\Omega_{i}$
and $G(t)=\sqrt{\nicefrac{2}{\pi}}\tau_{\mathrm{p}}\exp(-2t^{2}/\tau_{\mathrm{p}}^{2})$.
$\tau_{\mathrm{p}}$ is the effective pump-probe pulse cross-correlation
width,  $A_{\mathrm{O}i}$, $\Omega_{i}$, $\gamma_{i}$ are oscillator
amplitudes, frequencies and damping factors, respectively,  while
$A_{\mathrm{e}j}$ and $\tau_{j}$ are amplitudes and relaxation times
of the overdamped exponentially relaxing modes. 

We also assume that all oscillators are driven by a single exponential
term with relaxation time $\tau_{\mathrm{displ}}$, while the remaining
exponential relaxations are kept independent from the oscillators.
Note that the oscillator coordinates contain both the exponentially
relaxing and oscillatory components.

\begin{figure}
\includegraphics[angle=270,width=0.9\columnwidth]{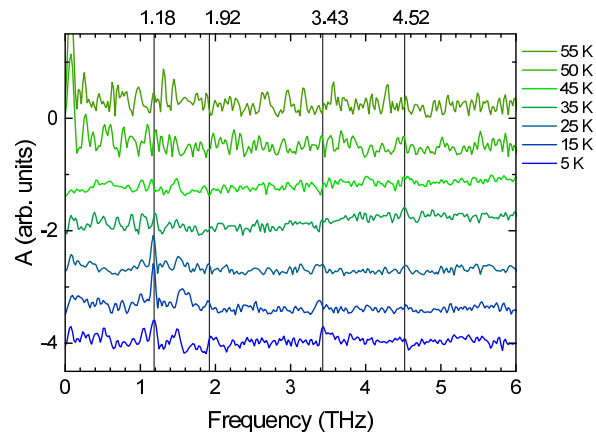}
\caption{Fourier transforms of the $T-$dependent data in Fig. \ref{fig:DRvsTlow}
with the DCE-model fits subtracted. The traces are vertically shifted
and the main remaining coherent oscillation peaks are indicated by
vertical lines.}
\label{fig:FFTvsTlow} 
\end{figure}

\begin{figure}
\includegraphics[angle=270,width=0.9\columnwidth]{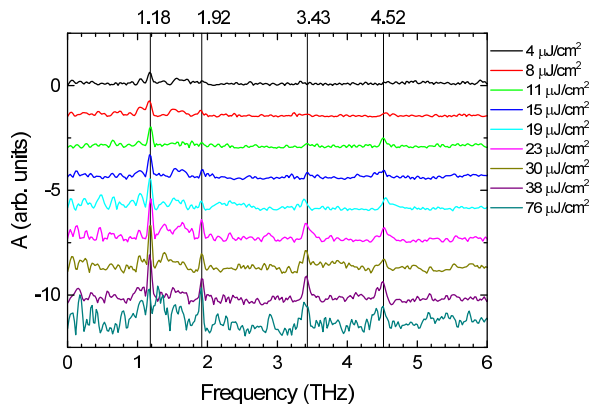}
\caption{Fourier transforms of the fluence dependent data in Fig. \ref{fig:DRvsF}
with the DCE-model fits subtracted. The traces are vertically shifted
and the main remaining coherent oscillation peaks are indicated by
vertical lines.}
\label{fig:Fourier} 
\end{figure}

It turns out that it is possible to fit the main features of the data
using four oscillators at $\sim1$ , $\sim0.8$, $\sim0.7$ and $\sim0.5$
THz with two independent exponential relaxations ($\tau_{1}\sim20$
ps and $\tau_{2}\sim10$ ns) in addition to the main relaxation component
($\tau_{\mathrm{displ}}\lesssim1$ ps) that drives the oscillators.
Fourier transforms of the fit residua shown in Figs. \ref{fig:FFTvsTlow}
and \ref{fig:Fourier} reveal additional four clearly resolvable modes
at the lowest $T$.

\section{Discussion}

\begin{figure}
\includegraphics[width=1\columnwidth]{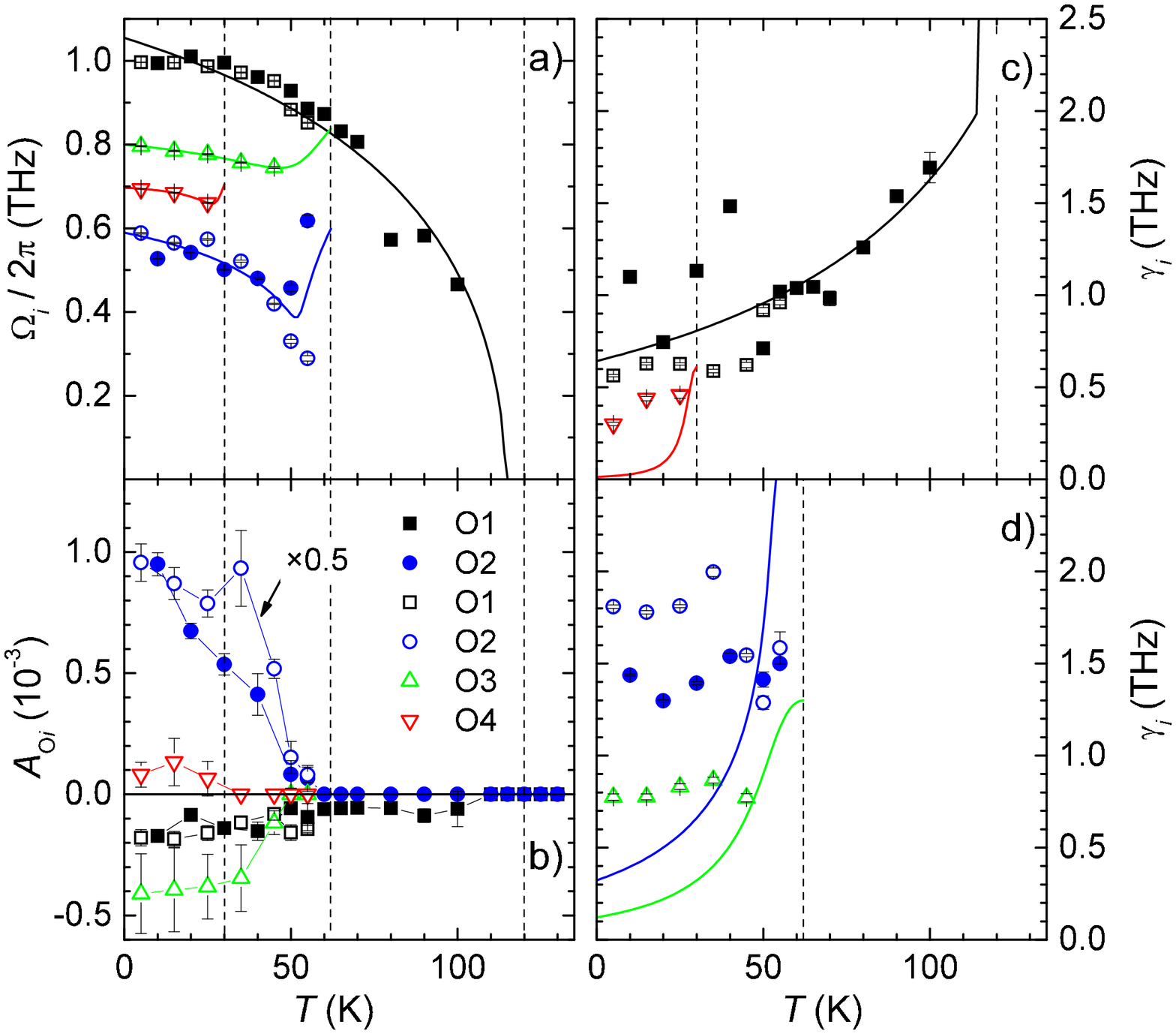} \caption{Temperature dependence of the oscillatory-modes DEC-model fit parameters.
The continuous lines in (a), (c) and (d) are the TDGL theory\cite{Schaefer2010,schaefer2014collective}
fits. Full and open symbols represent data from two separate experimental
runs from Figs. \ref{fig:Figure3} and \ref{fig:DRvsTlow}, respectively.
The vertical dashed lines correspond to the three CDW transition temperatures.}
\label{fig:DECP_T_O}
\end{figure}

The presence of collective coherent oscillatory excitations in the
photoinduced transient response of various CDWs is ubiquitous \cite{DemsarBlueBronze,Shimtake2007,YusupovPRL101,Schmitt2008,Perfetti2008,TomeljakPRL}
and comes from the coupling of phonons to the electronic order parameter
which drives the transition\cite{Schaefer2010}. In most cases several
weakly damped coherent oscillations are observed well below the CDW
transition temperatures, with widely different degrees of softening
when approaching the transition from below.\cite{DemsarBlueBronze,DemsarPRB,TomeljakPRL,Shimtake2007,YusupovPRL101}
The softening has been parametrized using the time dependent Ginzburg-Landau
(TDGL) theory\cite{Schaefer2010,schaefer2014collective} by the adiabaticity
ratio: 
\begin{equation}
a=\kappa m^{2}/\Omega_{0}^{3},\label{eq:adiab}
\end{equation}
where $\kappa$ is the bare damping of the electronic mode, $\Omega_{0}$
the bare phonon frequency and $m$ the coupling between the electronic
order parameter and the bare phonon mode. $a\gg1$ corresponds to
the adiabatic limit corresponding to the strongest degree of phonon
softening. $a\sim1$ corresponds to the non-adiabatic case with no
softening and increased damping of the oscillatory component when
approaching the transition temperature from below. In this case the
soft mode is the critically damped solution dominated by the electronic
order parameter.\cite{schaefer2014collective}

In the case of (PO$_{2}$)$_{4}$(WO$_{3}$)$_{12}$ the main coherent
oscillations at $\sim1$ THz and $\sim0.5$ THz appear strongly damped
down to the lowest $T$ with significantly weaker less-damped modes
observed only below $T\mathrm{_{CDW3}}$. 

Despite being rather weak the $1$-THz mode (O1) could be associated
with the formation of CDW1 at $T\mathrm{_{CDW1}}$. When approaching
$T\mathrm{_{CDW1}}$ from below (see Fig. \ref{fig:DECP_T_O}) it
softens by $\sim50$\% and with an increase of damping vanishes above
$\sim100$ K. 

Below $T\mathrm{_{CDW2}}$ two additional modes appear. The strongest
is the $\sim0.5$-THz mode (O2) which shows $\sim30\%$ softening
with increasing $T$ and vanishes above 55 K. Above 50 K the frequency
shows large scatter between the two experimental runs, presumably
due to a small amplitude and rather large damping that couples the
fit parameters of modes O1 and O2. The second mode at $\sim0.8$ Thz
(O3) is weaker, but less damped with a softening of less than $10$\%
disappearing above $T\sim50$ K. 

Below $T\mathrm{_{CDW3}}$ another two modes appear. The strongest
$\sim0.7$ THz mode (O4) shows $\sim5$\% softening with increasing
$T$, while the weaker one at $\sim1.2$ Thz, that was not included
in fits (see Figs. \ref{fig:FFTvsTlow} and \ref{fig:Fourier}) shows
virtually no softening. 

The three higher frequency modes observed clearly at higher fluences
(see Fig. \ref{fig:Fourier}) can not be clearly associated with any
of the transitions due to the weak intensities at the pump fluence
used for the $T$ scans.

\begin{figure}
\includegraphics[width=0.55\columnwidth]{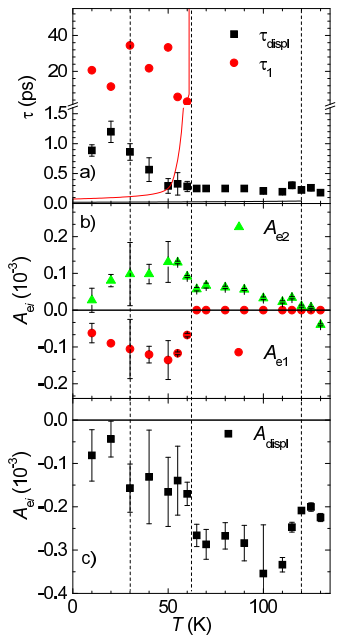} \caption{Temperature dependence of the exponential-modes DCE-model fit parameters
from Fig. \ref{fig:Figure3}. The red continuous line in (a) is the
TDGL theory\cite{Schaefer2010,schaefer2014collective} prediction
corresponding to the mode-O2 fit in Fig. \ref{fig:DECP_T_O}. The
vertical dashed lines correspond to the three CDW transition temperatures.}
\label{fig:DECP_T_A}
\end{figure}

\begin{table}
\begin{ruledtabular}%
\begin{tabular}{ccccc}
mode & $\Omega_{\mathrm{expt}}/2\pi$ (THz) & $\Omega_{0}/2\pi$ (THz) & $a$ & $T\mathrm{_{CDW}}$ (K)\tabularnewline
\hline 
O1 & 1.0 & 1.6 & 2.8 & 120\tabularnewline
O2 & 0.53 & 0.75 & 1.2 & 62\tabularnewline
O3 & 0.79 & 0.79 & 0.3 & 62\tabularnewline
O4 & 0.69 & 0.71 & 0.3 & 30\tabularnewline
\end{tabular}

\end{ruledtabular}\caption{The TDGL fit parameters for the four main coherent oscillatory modes
obtained from the fits in Fig. \ref{fig:DECP_T_O}.}
\label{tbl:TDGL}
\end{table}

\begin{figure}
\includegraphics[width=0.5\columnwidth]{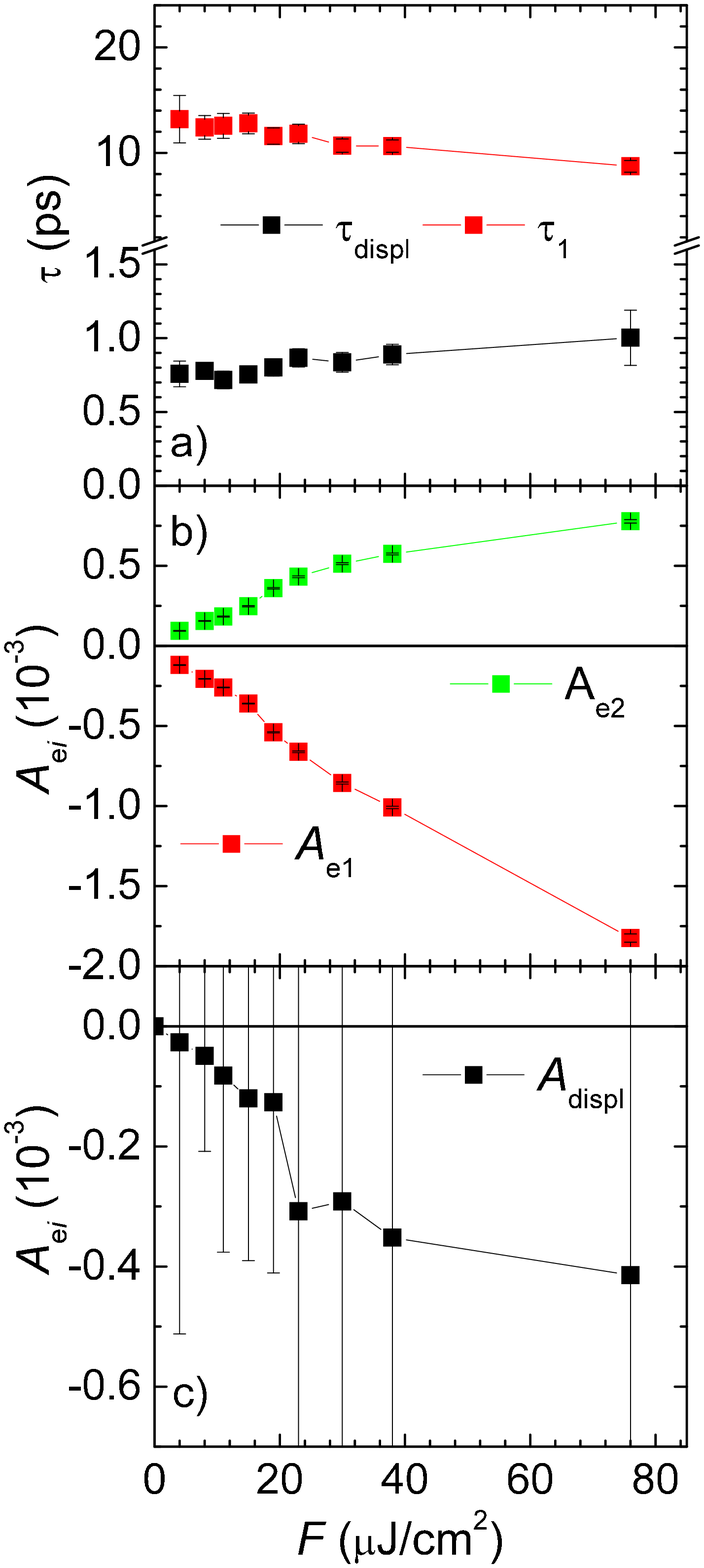}\includegraphics[width=0.5\columnwidth]{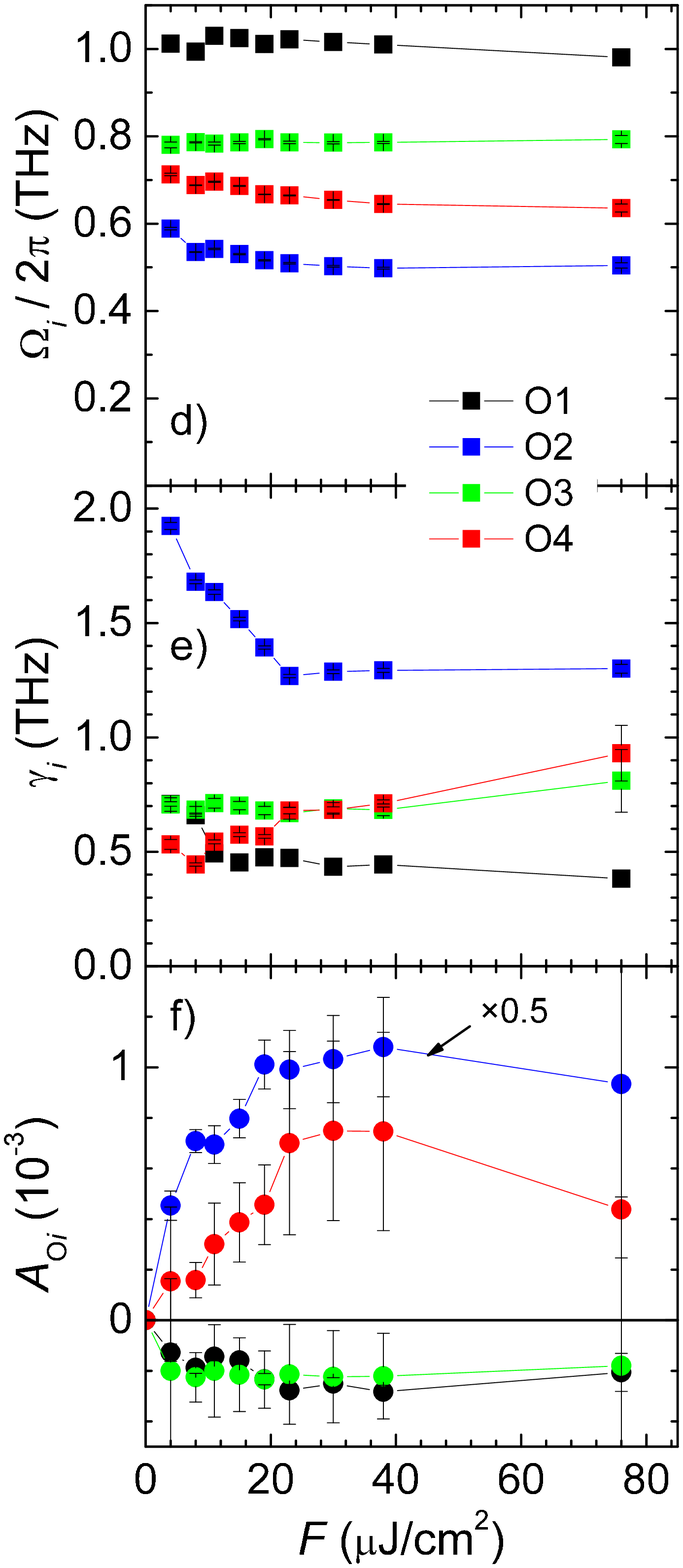}

\caption{Fluence dependence of the DCE-model fit parameters at $T=5$ K from
Fig. \ref{fig:DRvsF}. The large error bars in (c) are due to the
small amplitude of the displacive component in comparison to the oscilator
components amplitudes.}
\label{fig:DECP_F}
\end{figure}

The $T$ dependence of the mode frequencies is consistent with predictions
of the TDGL model\cite{Schaefer2010,schaefer2014collective} as shown
in Fig. \ref{fig:DECP_T_O} (a). However, except for mode O1 {[}see
Fig. \ref{fig:DECP_T_O}(c) and (d){]} it is not possible to obtain
simultaneous good fits for damping. The experimental dampings greatly
exceed the model predictions for modes O2, O3 and O4, except near
the CDW transitions, where the model dampings increase. We attribute
excess damping at lower $T$ to the combination of extrinsic and intrinsic
phonon damping that is not directly related to the CDWs and is not
included in the TDGL model\cite{Schaefer2010,schaefer2014collective}.
The model also predicts a region of critical damping for mode O1 just
below $T\mathrm{_{CDW1}}$ which is due to the vanishing amplitude
experimentally inaccessible. 

The main TDGL-theory parameters are shown in Table \ref{tbl:TDGL}.
The bare frequencies of modes O1 and O2 appear strongly renormalized
by the coupling to the electronic order parameter, while the renormalization
is small for the other modes.

The model\cite{Schaefer2010} also predicts a second critically damped
mode, that is weakly $T$-dependent in the adiabatic case and softens
at $T\mathrm{_{CDW}}$ in the non-adiabatic case. For CDW1 the mode-O1
fit in Fig. \ref{fig:DECP_T_O} (a) and (c) implies a fast virtually
$T$-independent mode with $\tau\sim20-50$ fs, which is ten times
faster that the fastest observed relaxation. If the excitation goes
through a displacive mechanism involving electrons in the ungapped
bands the fast mode would adiabatically follow the displacive drive
resulting in the response consistent with the data. 

In contrast, for CDW2 and CDW3 the fits to modes O2, O3 and O4 in
Fig. \ref{fig:DECP_T_O} (a) imply overdamped soft modes with rather
fast relaxation times diverging at the corresponding $T\mathrm{_{CDW}}$
as shown for mode O2 by the red solid line in Fig \ref{fig:DECP_T_A}
(a). None of the three observed exponential relaxation components
($\tau_{\mathrm{displ}}$, $\tau_{1}$, and $\tau_{2}$) show such
a divergence and therefore cannot be directly associated with the
overdamped part of the order parameter dynamics. This indicates that
photons at the energy used in the present experiment (1.55 eV) are
only weakly coupled to the electronic order parameters associated
with CDW2 and CDW3.

The present behaviour is different from observations\cite{PRBMilos}
in structurally similar\footnote{Both compounds contain similar octahedral layers separated by tetrahedral
units, but with different stacking.} $\eta$-Mo$_{4}$O$_{11}$ where a diverging relaxation time of the
overdamped soft mode is clearly observed while the phonon modes associated
with the CDW formation show virtually no softening when approaching
the CDW transition temperature. This is consistent with the the non-adiabatic
limit of the TDGL theory. One of the reasons for a smaller adiabaticity
ratio $a$ {[}see Eq. (\ref{eq:adiab}){]} could be higer frequencies
of the CDW related modes (1.93, 2.19 and 2.94 THz) in $\eta$-Mo$_{4}$O$_{11}$
originating in the smaller Mo mass. 

Looking at the pump fluence dependence of the DCE-model fit parameters
in Fig. \ref{fig:DECP_F} we observe a clear saturation of the coherent-oscillatory-mode
amplitudes above $\mathcal{F}_{\mathrm{sat}}\sim20$ $\mu$J/cm$^{2}$
while the amplitudes of the exponential relaxation components show
only a slightly sublinear behavior. Surprisingly, within error bars
the amplitudes of all the oscillatory modes\footnote{The amplitude of mode O3 appears saturated below the lowest measured
pump fluence. We believe that this is a fit artifact due to a strong
correlation of the fit amplitude with mode O4, which has a very similar
frequency, but the opposite phase.} saturate at the same fluence. Except for the damping of the strongest
mode O2 the other coherent-oscillatory-mode parameters also show no
strong changes around $\mathcal{F_{\mathrm{sat}}}$ and remain nearly
constant up to $\mathcal{F\sim}80$ $\mu$J/cm$^{2}$. 

Previously it has been shown that increasing the pump fluence transiently
suppresses and eventually destroys the CDW order.\cite{Perfetti2008,Schmitt2008,TomeljakPRL,YusupovNature}
Since in our experiment the CDW-induced modes remain clearly visible
up to $\mathcal{F\sim}80$ $\mu$J/cm$^{2}$ and are completely suppressed
only at higher fluences the saturation at $\mathcal{F_{\mathrm{sat}}}$
can not be attributed to the complete destruction of CDWs. It is however
unusual, that even the amplitude of mode O4 associated with the weakest
CDW3 shows only a moderate additional suppression with increasing
$\mathcal{F}$ above $\mathcal{F_{\mathrm{sat}}}$. 

A similar behavior in 1\emph{T}-TiSe$_{2}$, where the periodic lattice
distortion (PLD) persists upon destruction of the electronic ordering,
has been attributed\cite{porerLeierseder2014} to sequential suppression
of the electronic excitonic component of the CDW order at a lower
and the Jahn-Teller component at a higher pump fluence. In the case
of (PO$_{2}$)$_{4}$(WO$_{3}$)$_{12}$ it is believed that the CDWs
are well described in the framework of the Pierls-like mechanism with
\textquotedbl{}hidden\textquotedbl{} nesting vectors.\cite{FouryEPL,IJMPPouget93}
In this picture the PLD and the electronic part of the order parameter
work in conjunction to form CDWs. However, on a short timescale they
can become decoupled due to their different intrinsic timescales.
The saturation of the coherent-oscillatory-mode amplitudes can be
therefore attributed to such decoupling. In the case when the electronic
parts of CDWs are transiently completely suppressed and partially
recover on a timescale $\tau_{\mathrm{rec}}\ll\Omega_{i}^{-1}$ the
PLDs can not be completely suppressed. The rather short relaxation
times, $\tau_{i}\lesssim100$ fs, of the overdamped modes suggested
by the TDGL model fits are consistent with such scenario. 

Independently of the above saturation scenario the decrease of the
mode O2 damping with increasing fluence below $\mathcal{F_{\mathrm{sat}}}$
could be attributed to the inhomogeneous excitation profile. Due to
the finite light penetration depth the CDWs are more strongly suppressed
near the surface. The relative contribution of the near-surface part
of the excited volume to the response is therefore decreased leading
to a decrease of damping in the case of an enhanced extrinsic damping
of the mode due the surface effects.

\section{Summary and Conclusions}

We investigated the effect of a sequence of three consecutive CDW
transitions in a mono-phosphate tungsten bronze (PO$_{2}$)$_{4}$(WO$_{3}$)$_{2m}$
($m=6$) on the photoinduced ultrafast transient coherent oscillatory
optical response. We clearly observe the appearance of new coherent
oscillatory modes below each CDW transition. At low $T$ the interference
of two rather strongly damped modes at $\sim1$ and $\sim0.5$ THz,
which can be associated with CDW1 at 120 K and CDW2 at 62K, respectively,
leads to an unusual \emph{rectified} transient reflectivity with an
inverted-Gaussian shape. 

The $T$-dependence of the coherent mode frequencies can be well described
in the framework of the recently proposed TDGL model\cite{Schaefer2010,schaefer2014collective}.
The damping of the most strongly coupled mode at $\sim1$ THz, which
is coupled with CDW1 order parameter, is consistent with the TDGL
model. On the other hand, the less strongly coupled modes associated
with CDW2 and CDW3 show an additional intrinsic damping that does
not originate from the coupling to the electronic order parameters. 

Contrary to the oscillatory modes, no corresponding overdamped modes
predicted by the TDGL model are observed using probe photons at 1.5
eV, which may be explained to be due to weak coupling between the
1.55-eV photons and the electronic parts of the respective order parameters.
Further broad-band-probe transient reflectivity studies would be necessary
to possibly reveal the overdamped modes.

With increasing the pump fluence we observe a saturation of the coherent
mode amplitudes well below the complete destruction of the CDWs. This
could be attributed to a decoupling of the electronic and lattice
parts of the order parameter resulting in a complete ultrafast transient
suppression and a partial recovery of the electronic part of the order
parameter on a timescale much shorter than the coherent-mode periods.

\section{Acknowledgments}

The authors acknowledge Slovenian Ministry of education, science and
sport (project ULTRA-MEM-DEVICE No. PR-05665), Slovenian Research
Agency and European Research Council advanced grant TRAJECTORY for
financial support. We acknowledge D. Groult and Ph. Labbé at CRISMAT
(Caen, France) for providing the samples. We would like to thank also
P. Gubeljak for helping with the optical measurements and V.V. Kabanov
and J. Demsar for fruitful discussions.

\bibliographystyle{apsrev4-1}
\bibliography{biblio}

\end{document}